\documentclass[a4paper]{article}
\usepackage{authblk}
\usepackage[bookmarksopen]{hyperref}
\usepackage{amsmath}
\usepackage{amsfonts}
\usepackage{mathtools}
\usepackage{subcaption}
\usepackage{graphicx}
\usepackage{slashed}
\usepackage{braket}
\usepackage{ifthen}
\usepackage{tikz}
\usetikzlibrary{decorations.pathmorphing}
\usetikzlibrary{decorations.markings}
\usetikzlibrary{calc}
\usetikzlibrary{arrows}

\title{
 On $Z \to \gamma \gamma$ decay and cancellation of axial anomaly in
 $Z \to \gamma \gamma$ transition amplitude for massive fermions
}

\author{E.V. Zhemchugov\footnote{jini.zh@gmail.com}}
\affil{Institute for Theoretical and Experimental Physics, 117218, Moscow,
Russia}

\date{}

\numberwithin{equation}{section}

\newcommand{\zgg}{$Z \to \gamma \gamma$}
\renewcommand{\l}{\left}
\renewcommand{\r}{\right}
\newcommand{\s}{\slashed}
\renewcommand{\d}{\mathrm{d}}
\newcommand{\p}{\partial}
\renewcommand{\L}{\mathcal{L}}
\newcommand{\e}{\mathrm{e}}

\newcommand{\Int}{\int\limits}
\newcommand{\Sum}{\sum\limits}

\newcommand{\abs}[1]{\lvert #1 \rvert}

\newcommand{\row}[2]{\begin{pmatrix} #1 & #2 \end{pmatrix}}
\newcommand{\column}[2]{\begin{pmatrix} #1 \\ #2 \end{pmatrix}}

\newcommand{\xcoupling}[4]{\l[ #1 #2 #3 \r]\ifthenelse{\equal{#4}{}}{}{_{#4}}}
\newcommand{\coupling}[2]{\xcoupling{\bar #1}{#2}{#1}{}}
\newcommand{\vcoupling}[2]{\xcoupling{\bar #1}{#2}{#1}{V}}
\newcommand{\acoupling}[2]{\xcoupling{\bar #1}{#2}{#1}{A}}
\newcommand{\bcoupling}[2]{\l[ #1 #2 \r]}

\DeclareMathOperator{\tr}{tr}
\DeclareMathOperator{\arccot}{arccot}

\begin{document}

\tikzset{
 scalar/.style = {dashed},
 boson/.style = {
  solid,
  decorate,
  decoration = {snake}
 },
 fermion/.style = {
  solid,
  postaction = {
   decorate,
   decoration = {markings, mark = at position 0.55 with {\arrow{>}}}}
  }
}

\maketitle

\begin{abstract}
$Z \to \gamma \gamma$ decay amplitude is considered and proven to be zero due to
properties of polarization vectors and Bose statistics. Triangular diagrams for
a pseudoscalar~$\to \gamma \gamma$ and $Z \to \gamma \gamma$ processes with
massive fermions in the loop are explicitely calculated. In the Standard Model
axial anomaly vanishes in the sum of these diagrams as Z~boson is mixed with one
of the Goldstone bosons.

\end{abstract}

\section{Introduction}

A couple of articles has appeared recently concerning decay of a spin-1 particle
into two photons in spite of the fact that such a decay is prohibited by
Landau-Yang theorem~\cite{landau, yang}. Paper~\cite{zgg}, released in two
versions, one in 2011, and the other in 2013, considers $Z \to \gamma \gamma$
decay. The first version claimed that the decay width is not zero, albeit less
than the currently established experimental boundary. The decay width was
calculated from an amplitude which does not satisfy Ward identities for photons,
apparently by applying photon polarization matrix proportional to $g_{\mu \nu}$.
Since actual photon polarization matrix contains an ambiguous term proportional
to photon momentum which gives zero only as long as Ward identities are
satisfied, this calculation produced an incorrect result. It has been corrected
in the second version of paper~\cite{zgg}, by noticing that the amplitude equals
zero.  However, the expression for the amplitude has not been changed, so Ward
identities remained invalid. A proper expression for the amplitude both on and
off shell is provided in section~\ref{zgg} of this work.

Another paper,~\cite{higgs-1}, alerts us that we should keep considering the
126~GeV Higgs boson candidate as a particle with spin 1 in spite of the clearly
observed two photons decay of it. The paper claims that Landau-Yang theorem is
inapplicable in this case. While not addressing this statement, one can
demonstrate that decay of a spin-1 particle into two photons is impossible
considering only the tensor structure of the amplitude, Bose statistics, and
properties of polarization vectors of photons and the decaying particle. This
demonstration is provided in section~\ref{zgg-intro}.

$Z \to \gamma \gamma$ transition amplitude is a textbook example of an axial
anomaly appearing in a triangle diagram. A self-consistent theory has to be free
of anomalies. In the Glashow-Weinberg-Salam theory of weak interactions with
massless fermions, due to $U(1)$ hypercharge values, anomalies cancel out when a
sum of all possible fermions running in the fermion loop is accounted
for~\cite{bouchiat}. However, in the case of massive fermions, straightforward
calculation of $Z \to \gamma \gamma$ amplitude shows that its derivative is
proportional to a term dependent on fermion mass.  Since fermion masses are free
parameters of the theory, contributions of different fermions in the loop can no
longer cancel out. Nevertheless, the Standard Model features a mechanism to keep
anomalies being zero which stems from the way fermion masses are
generated---spontaneous symmetry breaking. In section~\ref{mgws} it is shown
that Z boson is mixed with one of the Goldstone bosons, and the latter provides
the exact value to cancel out the mass-dependent term of the derivative of $Z
\to \gamma \gamma$ amplitude. Since mass-independent terms keep cancelling out
in the same way as in the massless theory, it is concluded that in the Standard
Model $Z \to \gamma \gamma$ transition amplitude is free of anomalies.

\section{$Z \to \gamma \gamma$ decay}

\label{zgg-intro}

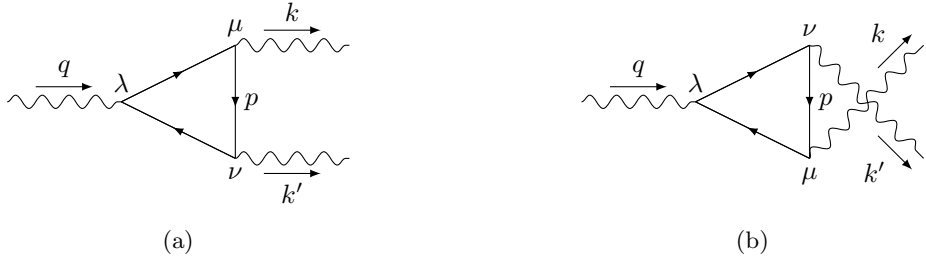
\begin{figure}[h]
 \centering
 \begin{subfigure}{0.4\textwidth}
  \centering

  \begin{tikzpicture}[>=latex]
   \begin{scope}[shape=coordinate]
    \node (q) {}
     [grow=right]
     child {
      node [label={[shape=rectangle]above:$\lambda$}] (lambda) {}
      child { node [label={[shape=rectangle]below:$\nu$}] (nu) {}
              child {node (k') {} edge from parent [boson]}}
      child { node [label={[shape=rectangle]above:$\mu$}] (mu) {}
              child {node (k)  {} edge from parent [boson]}}
      edge from parent [boson]
     };
   \end{scope}

   \draw[fermion,auto] (lambda) -- (mu);
   \draw[fermion,auto] (mu) to node {$p$} (nu);
   \draw[fermion,auto] (nu) -- (lambda);

   \draw[->,auto] ([yshift=2mm]$(q)!0.25!(lambda)$)
                  to node {$q$}
                  ([yshift=2mm]$(q)!0.75!(lambda)$);
   \draw[->,auto] ([yshift=2mm]$(mu)!0.25!(k)$)
                  to node {$k$}
                  ([yshift=2mm]$(mu)!0.75!(k)$);
   \draw[->,auto] ([yshift=-2mm]$(nu)!0.25!(k')$)
                  to node [swap] {$k'$}
                  ([yshift=-2mm]$(nu)!0.75!(k')$);

   \coordinate (diagramne) at (current bounding box.north east);
   \coordinate (diagramsw) at (current bounding box.south west);
  \end{tikzpicture}
  \caption{ }

  \label{diag:z-gg-a}
 \end{subfigure}
 ~
 \begin{subfigure}{0.4\textwidth}
  \centering

  \begin{tikzpicture}[>=latex]
   \useasboundingbox (diagramne) rectangle (diagramsw);

   \begin{scope}[shape=coordinate]
    \node (q) {}
     [grow=right]
     child {
      node [label={[shape=rectangle]above:$\lambda$}] (lambda) {}
      child {
       node [label={[shape=rectangle]below:$\mu$}] (mu) {}
       child {node (k')  {} edge from parent [draw=none]}
      }
      child {
       node [label={[shape=rectangle]above:$\nu$}] (nu) {}
       child {node (k) {} edge from parent [draw=none]}
      }
      edge from parent [boson]
     };
   \end{scope}

   \draw[boson] (mu) -- (k);
   \draw[boson] (nu) -- (k');

   \draw[fermion,auto] (lambda) -- (nu);
   \draw[fermion,auto] (nu) to node {$p$} (mu);
   \draw[fermion,auto] (mu) -- (lambda);

   \draw[->,auto] ([yshift=2mm]$(q)!0.25!(lambda)$)
                  to node {$q$}
                  ([yshift=2mm]$(q)!0.75!(lambda)$);
   \draw[->,auto] ([yshift=3mm]$(mu)!0.6!(k)$)
                  to node {$k$}
                  ([yshift=3mm]$(mu)!0.9!(k)$);
   \draw[->,auto] ([yshift=-3mm]$(nu)!0.6!(k')$)
                  to node [swap] {$k'$}
                  ([yshift=-3mm]$(nu)!0.9!(k')$);
  \end{tikzpicture}

  \caption{ }
  \label{diag:z-gg-b}
 \end{subfigure}
 \caption{Lowest-order Feynman diagrams of \zgg{} transition.}
 \label{diag:z-gg}
\end{figure}

Amplitude for $Z \to \gamma \gamma$ decay $\mathcal{M}_{Z \to \gamma \gamma}$ is
proportional to the following expression:
\begin{equation}
 \mathcal{M}_{Z \to \gamma \gamma}
 \sim \epsilon_\mu \epsilon'_\nu \tilde \epsilon_\lambda
      T^{\mu \nu \lambda}(k, k'),
 \label{z-gg-intro:amplitude}
\end{equation}
where $\epsilon$ and $\epsilon'$ are polarization vectors of photons, $\tilde
\epsilon$ is the polarization vector of the Z~boson, $k$ and $k'$ are photons
momenta, and $T^{\mu \nu \lambda}(k, k')$ is the tensor corresponding to the sum
of diagrams~\ref{diag:z-gg}:
\begin{equation}
 T^{\mu \nu \lambda}(k, k')
 = \tilde T^{\mu \nu \lambda}(k, k') + \tilde T^{\nu \mu \lambda}(k', k),
 \label{z-gg-intro:T-decomposition1}
\end{equation}
where $\tilde T^{\mu \nu \lambda}(k, k')$ is the tensor corresponding to
diagram~\ref{diag:z-gg-a}.

Z~boson couples with fermions via both vector and axial currents of the form
\begin{equation}
 (g_V \bar \psi \gamma^\mu \psi + g_A \bar \psi \gamma^\mu \gamma^5 \psi)
 Z_\mu,
\end{equation}
where $g_V$ and $g_A$ are coupling constants. It follows from the Furry's
theorem~\cite[\textsection 79]{landafshitz}, that proportional to the vector
coupling part of the amplitude equals zero.  Thus, we need to consider only
axial coupling. The most general representation of the latter
is~\cite{rosenberg}%
\footnote{
 Note the absence of the $k^\lambda \varepsilon^{\mu \nu \alpha \beta} k_\alpha
 k'_\beta$ term.  This is due to it being linearly dependent on other terms as
 follows from~\eqref{tr7:identity}.
}
\begin{equation}
 \tilde T^{\mu \nu \lambda}(k, k')
 = (A k_\alpha + A' k'_\alpha) \varepsilon^{\mu \nu \lambda \alpha}
 + (M k^\mu   + M' k'^\mu)
   \varepsilon^{\nu \lambda \alpha \beta} k'_\alpha k_\beta
 + (N k^\nu    + N' k'^\nu)
   \varepsilon^{\mu \lambda \alpha \beta} k'_\alpha k_\beta
 \label{z-gg-intro:T-decomposition2}
\end{equation}
where $A$, $A'$, $M$, $M'$, $N$, $N'$ are some functions depending on $k$
and $k'$. Substitution of~\eqref{z-gg-intro:T-decomposition2}
into~\eqref{z-gg-intro:T-decomposition1} results in the following expression:
\begin{equation}
 \begin{split}
   T^{\mu \nu \lambda}(k, k')
  &= ((A - A'^*) k_\alpha + (A' - A^*) k'_\alpha)
     \varepsilon^{\mu \nu \lambda \alpha}
  \\
  &+ ((N - M'^*) k^\nu + (N' - M^*) k'^\nu)
     \varepsilon^{\mu \lambda \alpha \beta} k'_\alpha k_\beta
  \\
  &+ ((M - N'^*) k^\mu + (M' - N^*) k'^\mu)
     \varepsilon^{\nu \lambda \alpha \beta} k'_\alpha k_\beta,
 \end{split}
 \label{z-gg-intro:T-tensor}
\end{equation}
where $A^*$, $A'^*$, \ldots{} are $A$, $A'$, \ldots{} with $k$ and $k'$
interchanged.

The expression for the amplitude should satisfy Ward identities
\begin{equation}
 k_\mu T^{\mu \nu \lambda}(k, k') = 0,
 \quad
 k'_\nu T^{\mu \nu \lambda}(k, k') = 0.
 \label{z-gg-intro:ward}
\end{equation}
The third Ward identity,
\begin{equation}
 q_\lambda T^{\mu \nu \lambda}(k, k') = 0,
 \label{z-gg-intro:ward3}
\end{equation}
where $q = k + k'$ is Z boson momentum, is violated, and this fact is referred to
as axial anomaly. We will consider it in section~\ref{gws}.
Eqs.~\eqref{z-gg-intro:T-tensor},~\eqref{z-gg-intro:ward} provide the following
relation for the coefficients:
\begin{equation}
 A' - A^* = (M - N'^*) k^2 + (M' - N^*) kk',
\end{equation}
and another one with $k$ and $k'$ interchanged. Hence,
$T^{\mu \nu \lambda}(k, k')$ has the following structure:
\begin{equation}
 \begin{split}
  T^{\mu \nu \lambda}(k, k')
  &= (N' - M^*)
     (  k'^2   \varepsilon^{\mu \nu \lambda \alpha} k_\alpha
      + k'^\nu \varepsilon^{\mu \lambda \alpha \beta} k'_\alpha k_\beta)
   + (N - M'^*)
     (  kk'   \varepsilon^{\mu \nu \lambda \alpha} k_\alpha
      + k^\nu \varepsilon^{\mu \lambda \beta \alpha} k'_\alpha k_\beta)
  \\
  &+ (M - N'^*)
     (  k^2   \varepsilon^{\mu \nu \lambda \alpha} k'_\alpha
      + k^\mu \varepsilon^{\nu \lambda \alpha \beta} k'_\alpha k_\beta)
   + (M' - N^*)
     (  kk'    \varepsilon^{\mu \nu \lambda \alpha} k'_\alpha
      + k'^\mu \varepsilon^{\nu \lambda \alpha \beta} k'_\alpha k_\beta).
 \end{split}
\end{equation}
On the mass shell $k^2 = k'^2 = 0$, and there is only one Lorentz invariant
scalar which depends on either $k$ or $k'$: $kk'$. It does not change under $k
\leftrightarrow k'$ interchange, consequently, on the mass shell $M = M^*$, $M'
= M'^*$, \ldots. Therefore,
\begin{equation}
 T^{\mu \nu \lambda}(k, k')
 = (N - M')
   \Bigl( 
      kk' \varepsilon^{\mu \nu \lambda \alpha} (k - k')_\alpha
    + (  k^\nu  \varepsilon^{\mu \lambda \alpha \beta}
       - k'^\mu \varepsilon^{\nu \lambda \alpha \beta})
      k'_\alpha k_\beta
   \Bigr)
 + (N' - M)
   (
      k'^\nu \varepsilon^{\mu \lambda \alpha \beta}
    - k^\mu  \varepsilon^{\nu \lambda \alpha \beta}
   )
   k'_\alpha k_\beta.
 \label{z-gg-intro:T-onshell}
\end{equation}

Let us now work in a system of coordinates where Z~boson is at rest and consider
relations between vectors appearing in the problem. Let the $z$ axis be parallel
to the spatial part of the photon momentum $k$. Then photon momenta can be
written as follows:
\begin{equation}
 k = (k_0, 0, 0, k_0), \quad k' = (k_0, 0, 0, -k_0),
\end{equation}
where $4 k_0^2 = 2 kk' = (k + k')^2 = q^2 = m_Z^2$ is Z~boson mass squared.
Photon polarization vectors are orthogonal to photon momenta, and can be chosen
as follows:
\begin{equation}
 \epsilon_\mu = (0, \epsilon_1, \epsilon_2, 0),
 \quad
 \epsilon'_\nu = (0, \epsilon'_1, \epsilon'_2, 0).
\end{equation}
Finally, let us take the physical polarization of Z~boson:
\begin{equation}
 \tilde \epsilon_\lambda
 = (0, \tilde \epsilon_1, \tilde \epsilon_2, \tilde \epsilon_3).
\end{equation}
With these equations in mind it becomes clear that substitution
of~\eqref{z-gg-intro:T-onshell} into~\eqref{z-gg-intro:amplitude} gives zero,
because:
\begin{enumerate}
 \item $\epsilon_\mu k^\mu = \epsilon'_\nu k'^\nu = 0$,
 \item $\epsilon_\mu k'^\mu = \epsilon'_\nu k^\nu = 0$,
 \item $
  \epsilon_\mu \epsilon'_\nu \tilde \epsilon_\lambda
  \varepsilon^{\mu \nu \lambda \alpha} (k - k')_\alpha
  = 0
 $ since the only $\alpha$ when $
  \epsilon_\mu \epsilon'_\nu \tilde \epsilon_\lambda
  \varepsilon^{\mu \nu \lambda \alpha} \ne 0
 $ is $\alpha = 0$, but $(k - k')_0 = k_0 - k_0 = 0$.
\end{enumerate}
Hence, amplitude of \zgg{} decay is equal to zero in the Z~boson rest frame.
Since the amplitude is Lorentz invariant, it equals zero in any other coordinate
system as well, and \zgg{} decay amplitude vanishes.

It should be stressed that the amplitude vanishes on the mass shell; it does not
if one of the photons is off shell. For example, amplitude of the $Z \to \gamma
\gamma^* \to \gamma e^+ e^-$ transition gives nonzero contribution to the $Z \to
\gamma e^+ e^-$ decay amplitude.

\section{Triangle diagrams}

In section~\ref{mgws} we will need explicit expressions of two transition
amplitudes: \zgg{} and $\varphi \to \gamma \gamma$, where $\varphi$ is a
pseudoscalar particle. Calculation of these amplitudes is relatively simple but
tedious, and most of the necessary information is provided in any general
quantum field theory course, so we will omit some intermediate steps and provide
only the final results.

\subsection{$\varphi \to \gamma \gamma$}

\begin{figure}[h]
 \centering
 \begin{subfigure}{0.4\textwidth}
  \centering

  \begin{tikzpicture}[>=latex]
   \begin{scope}[shape=coordinate]
    \node (q) {}
     [grow=right]
     child {
      node (lambda) [label={[shape=rectangle]above:$\lambda$}] {}
      child { node [label={[shape=rectangle]below:$\nu$}] (nu) {}
              child {node (k') {} edge from parent [boson]}}
      child { node [label={[shape=rectangle]above:$\mu$}] (mu) {}
              child {node (k)  {} edge from parent [boson]}}
      edge from parent [scalar]
     };
   \end{scope}

   \draw[fermion,auto] (lambda) -- (mu);
   \draw[fermion,auto] (mu) to node {$p$} (nu);
   \draw[fermion,auto] (nu) -- (lambda);

   \draw[->,auto] ([yshift=2mm]$(q)!0.25!(lambda)$)
                  to node {$q$}
                  ([yshift=2mm]$(q)!0.75!(lambda)$);
   \draw[->,auto] ([yshift=2mm]$(mu)!0.25!(k)$)
                  to node {$k$}
                  ([yshift=2mm]$(mu)!0.75!(k)$);
   \draw[->,auto] ([yshift=-2mm]$(nu)!0.25!(k')$)
                  to node [swap] {$k'$}
                  ([yshift=-2mm]$(nu)!0.75!(k')$);
   \coordinate (diagramne) at (current bounding box.north east);
   \coordinate (diagramsw) at (current bounding box.south west);
  \end{tikzpicture}

  \caption{ }
  \label{diag:phi-gg-a}
 \end{subfigure}
 ~
 \begin{subfigure}{0.4\textwidth}
  \centering

  \begin{tikzpicture}[>=latex]
   \useasboundingbox (diagramne) rectangle (diagramsw);
   \begin{scope}[shape=coordinate]
    \node (q) {}
     [grow=right]
     child {
      node [label={[shape=rectangle]above:$\lambda$}] (lambda) {}
      child {
       node [label={[shape=rectangle]below:$\mu$}] (mu) {}
       child {node (k')  {} edge from parent [draw=none]}
      }
      child {
       node [label={[shape=rectangle]above:$\nu$}] (nu) {}
       child {node (k) {} edge from parent [draw=none]}
      }
      edge from parent [scalar]
     };
   \end{scope}

   \draw[boson] (mu) -- (k);
   \draw[boson] (nu) -- (k');

   \draw[fermion,auto] (lambda) -- (nu);
   \draw[fermion,auto] (nu) to node {$p$} (mu);
   \draw[fermion,auto] (mu) -- (lambda);

   \draw[->,auto] ([yshift=2mm]$(q)!0.25!(lambda)$)
                  to node {$q$}
                  ([yshift=2mm]$(q)!0.75!(lambda)$);
   \draw[->,auto] ([yshift=3mm]$(mu)!0.6!(k)$)
                  to node {$k$}
                  ([yshift=3mm]$(mu)!0.9!(k)$);
   \draw[->,auto] ([yshift=-3mm]$(nu)!0.6!(k')$)
                  to node [swap] {$k'$}
                  ([yshift=-3mm]$(nu)!0.9!(k')$);
  \end{tikzpicture}

  \caption{ }
  \label{diag:phi-gg-b}
 \end{subfigure}
 \caption{
  Lowest-order Feynman diagrams of $\varphi \to \gamma \gamma$ transition.
 }
 \label{diag:phi-gg}
\end{figure}
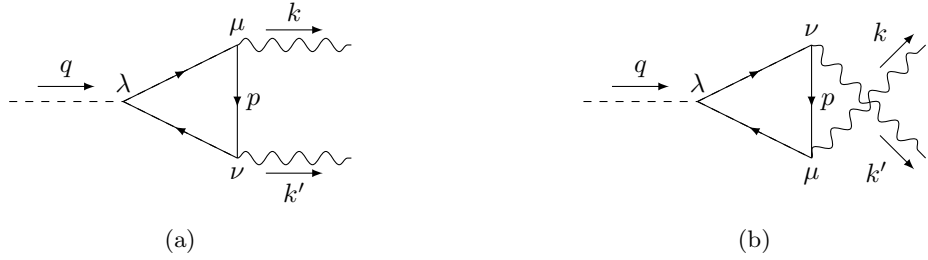

$\varphi \to \gamma \gamma$ transition amplitude is equal to the sum of Feynman
diagrams~\ref{diag:phi-gg-a} and~\ref{diag:phi-gg-b}:
\begin{equation}
 \begin{split}
   \mathcal{M}_{\varphi \to \gamma \gamma}(k, k')
  &= \int
      -\tr \l(
        \frac{i}{\s p - \s k' - m}
        i \coupling{\psi}{\gamma} \s \epsilon'
        \frac{i}{\s p - m}
        i \coupling{\psi}{\gamma} \s \epsilon
        \frac{i}{\s p + \s k - m}
        i \acoupling{\psi}{\varphi} \gamma^5
      \r)
     \frac{\d^4 p}{(2 \pi)^4}
   + (\epsilon \leftrightarrow \epsilon', k \leftrightarrow k')
  \\
  &= \coupling{\psi}{\gamma}^2 \acoupling{\psi}{\varphi}
     \epsilon_\mu \epsilon'_\nu T^{\mu \nu}(k, k'),
 \end{split}
 \label{phi-gg:phigg}
\end{equation}
where $\coupling{\psi}{\gamma}$ is the coupling constant between fermion and
photon (fermion electric charge), $\acoupling{\psi}{\varphi}$ is the
pseudoscalar coupling constant between $\varphi$ and fermion, $\epsilon$ and
$\epsilon'$ are photon polarization vectors, $m$ is the fermion mass,
\begin{gather}
 T^{\mu \nu}(k, k') = \tilde T^{\mu \nu}(k, k') + \tilde T^{\nu \mu}(k', k),
 \label{phi-gg:T}
 \\
 \tilde T^{\mu \nu}(k, k')
 = \int
    \frac{
     \tr \bigl(
      (\s p - \s k' + m) \gamma^\nu
      (\s p         + m) \gamma^\mu
      (\s p + \s k  + m) \gamma^5
     \bigr)
    }{
     \bigl( (p - k')^2 - m^2 \bigr)
     \bigl( p^2        - m^2 \bigr)
     \bigl( (p + k )^2 - m^2 \bigr)
    }
  \frac{\d^4 p}{(2 \pi)^4}.
 \label{phi-gg:T-integral}
\end{gather}

The trace is equal to\footnote{We use the following conventions:
$\gamma^5 = -i \gamma^0 \gamma^1 \gamma^2 \gamma^3$, $\varepsilon^{0123} = +1$.}
\begin{equation}
 \tr \bigl(
  (\s p - \s k' + m) \gamma^\nu
  (\s p         + m) \gamma^\mu
  (\s p + \s k  + m) \gamma^5
 \bigr)
 = 4 i m \varepsilon^{\mu \nu \alpha \beta} k'_\alpha k_\beta.
\end{equation}
Calculating the integral and substituting the result into~\eqref{phi-gg:phigg},
we obtain:
\begin{equation}
 \begin{split}
   \mathcal{M}_{\varphi \to \gamma \gamma}
  &= \coupling{\psi}{\gamma}^2 \acoupling{\psi}{\varphi}
     \epsilon_\mu \epsilon'_\nu T^{\mu \nu}(k, k')
  \\
  &= -\frac{m}{2 \pi^2}
     \coupling{\psi}{\gamma}^2 \acoupling{\psi}{\varphi}
     \varepsilon^{\mu \nu \alpha \beta}
     \epsilon_\mu \epsilon'_\nu k'_\alpha k_\beta
     \Int_0^1 \Int_0^{1-x}
      \frac{\d z \, \d x}{2 xz kk' - m^2 + x (1 - x) k^2 + z (1 - z) k'^2}.
 \end{split}
 \label{phi-gg:amplitude-off-shell}
\end{equation}
On the mass shell
\begin{equation}
 \mathcal{M}_{\varphi \to \gamma \gamma}
 = -\frac{m}{2 \pi^2} \coupling{\psi}{\gamma}^2 \acoupling{\psi}{\varphi}
   \varepsilon^{\mu \nu \alpha \beta}
   \epsilon_\mu \epsilon'_\nu k'_\alpha k_\beta
   \Int_0^1 \Int_0^{1-x}
    \frac{\d z \, \d x}{2 xz kk' - m^2}.
\end{equation}
Let
\begin{equation}
 \eta_\varphi = \l( \frac{m}{m_\varphi} \r)^2,
 \label{phi-gg:eta}
\end{equation}
where $m_\varphi$ is the mass of $\varphi$. Then
\begin{equation}
 \mathcal{M}_{\varphi \to \gamma \gamma}
 = -\frac{\coupling{\psi}{\gamma}^2 \acoupling{\psi}{\varphi}}
         {(2 \pi)^2 m_\varphi}
   \varepsilon^{\mu \nu \alpha \beta}
   \epsilon_\mu \epsilon'_\nu k'_\alpha k_\beta
   \sqrt{\eta_\varphi}
   \ln^2 \l( 1 - \frac{1 + \sqrt{1 - 4 \eta_\varphi}}{2 \eta_\varphi} \r).
\end{equation}
This expression can be rewritten as follows:
\begin{multline}
 \mathcal{M}_{\varphi \to \gamma \gamma}
 = \frac{\coupling{\psi}{\gamma}^2 \acoupling{\psi}{\varphi}}
        {(2 \pi)^2 m_\varphi}
   \epsilon_\mu \epsilon'_\nu \varepsilon^{\mu \nu \alpha \beta}
   k'_\alpha k_\beta
   \sqrt{\eta_\varphi}
 \\ \cdot \l\{
    \begin{aligned}
     &  \l(
           \ln^2 \l(
            \frac{1 + \sqrt{1 - 4 \eta_\varphi}}{2 \eta_\varphi} - 1
           \r)
         + \pi^2
        \r)
        \e^{
         -2i
         \arccot \frac{
          \pi
         }{
          \ln \l( \frac{1 + \sqrt{1 - 4 \eta_\varphi}}{2 \eta_\varphi} - 1 \r)
         }
        },
     & &\eta_\varphi \le \frac{1}{4},
     \\& \l(
          \pi - \arctan \frac{\sqrt{4 \eta_\varphi - 1}}{1 - 2 \eta_\varphi}
         \r)^2,
     & \frac{1}{4} \le &\eta_\varphi \le \frac{1}{2},
     \\& \arctan^2 \frac{\sqrt{4 \eta_\varphi - 1}}{2 \eta_\varphi - 1},
     & &\eta_\varphi \ge \frac{1}{2}.
    \end{aligned}
   \r.
 \label{phi-gg:amplitude-on-shell}
\end{multline}

\subsection{\zgg}

\label{zgg}

\zgg{} transition amplitude is equal to the sum of Feynman
diagrams~\ref{diag:z-gg-a} and~\ref{diag:z-gg-b}:
\begin{equation}
 \begin{split}
   \mathcal{M}_{\mathrm{Z} \to \gamma \gamma}
  &= \int -\tr \l(
      \frac{i}{\s p - \s k' - m}
      i \coupling{\psi}{\gamma}     \s \epsilon'
      \frac{i}{\s p         - m}
      i \coupling{\psi}{\gamma}     \s \epsilon
      \frac{i}{\s p + \s k  - m}
      i \acoupling{\psi}{Z} \tilde {\s \epsilon}
      \gamma^5
     \r)
     \frac{\d^4 p}{(2 \pi)^4}
     + (\epsilon \leftrightarrow \epsilon', k \leftrightarrow k')
  \\
  &= \coupling{\psi}{\gamma}^2 \acoupling{\psi}{Z}
     \epsilon_\nu \epsilon'_\mu \tilde \epsilon_\lambda
     T^{\mu \nu \lambda}(k, k'),
 \end{split}
 \label{z-gg:zgg}
\end{equation}
where $\coupling{\psi}{\gamma}$ is the coupling constant between fermion and
photon, $\acoupling{\psi}{Z}$ is the axial coupling constant between Z~boson and
fermion, $\epsilon$, $\epsilon'$, $\tilde \epsilon$ are polarization vectors of
photons and Z~boson, $m$ is the fermion mass,
\begin{gather}
 T^{\mu \nu \lambda}(k, k')
 = \tilde T^{\mu \nu \lambda}(k, k') + \tilde T^{\nu \mu \lambda}(k', k),
 \label{z-gg:T}
 \\
 \tilde T^{\mu \nu \lambda}(k, k')
 = \int \frac{
    \tr \bigl(
     (\s p - \s k' + m) \gamma^\nu
     (\s p         + m) \gamma^\mu
     (\s p + \s k  + m) \gamma^\lambda \gamma^5
    \bigr)
   }{
    \bigl( (p - k')^2 - m^2 \bigr)
    \bigl( p^2        - m^2 \bigr)
    \bigl( (p + k )^2 - m^2 \bigr)
   }
   \frac{\d^4 p}{(2 \pi)^4}.
 \label{z-gg:tilde-T}
\end{gather}

The trace is evaluated with the help of~\eqref{tr7:tr7}. Unlike the case
of~\eqref{phi-gg:T-integral}, now the trace depends on integration variable $p$.
The resulting expression is carried through Feynman parameterization, and when
the integration with respect to $p$ is performed, it is convenient to eliminate
terms proportional to $k^\lambda$ and $k'^\lambda$ with the help
of~\eqref{tr7:identity}. The final result is
\begin{equation}
 \begin{split}
   \tilde T^{\mu \nu \lambda} (k, k')
  &= \frac{1}{4 \pi^2} \Int_0^1 \Int_0^{1-x}
     \frac{1}{2 xz kk' - m^2 + x (1-x) k^2 + z (1-z) k'^2} \cdot
 \\ & \qquad
       \cdot 
        \Bigl\{
           \varepsilon^{\mu \nu \lambda \alpha}
           \Bigl[
             \bigl(
              x^2 (1-x) k^2 - z (1-z+xz) k'^2 - 2 xz (1-x) kk' - (1-x) m^2
             \bigr)
             k_\alpha
 \\ & \qquad \qquad
            - \bigl(
               z^2 (1-z) k'^2 - x (1-x+xz) k^2 - 2 xz (1-z) kk' - (1-z) m^2
              \bigr) k'_\alpha
           \Bigr]
 \\ & \qquad
         + 2 \Bigl[
              \varepsilon^{\nu \lambda \alpha \beta} x ((1-x) k + z k')^\mu
            - \varepsilon^{\mu \lambda \alpha \beta} z (x k + (1-z) k')^\nu
           \Bigr] k'_\alpha k_\beta
        \Bigr\}
    \, \d z \, d x.
 \end{split}
 \label{z-gg:tilde-T-unregularized}
\end{equation}
Integration with respect to Feynman parameters $x$ and $z$ is performed over the
region
\begin{equation*}
 \{ x, z \colon x \ge 0, z \ge 0, x + z \le 1 \},
\end{equation*}
which is invariant
with respect to interchange $x \leftrightarrow z$. Consequently, integration
variables can be interchanged independently of integration limits with no effect
on the value of $\tilde T^{\mu \nu \lambda}(k, k')$. Using this fact it is easy
to see that
\begin{equation}
 \tilde T^{\mu \nu \lambda}(k, k') = \tilde T^{\nu \mu \lambda}(k', k),
\end{equation}
hence
\begin{equation}
 T^{\mu \nu \lambda}(k, k') = 2 \tilde T^{\mu \nu \lambda}(k, k').
 \label{z-gg:T-unregularized}
\end{equation}

Note that although~\eqref{z-gg:tilde-T} is superficially
divergent,~\eqref{z-gg:tilde-T-unregularized} is finite. Divergent terms
disappeared when integration with respect to Feynman parameters was performed.
Paper~\cite{zgg} argues that in this case there is no need to regularize the
amplitude. However, in order to extract divergent terms a shift over the
integration variable $p$ was necessary. This shift introduced an ambiguous
surface term, and now \eqref{z-gg:tilde-T-unregularized} does not satisfy Ward
identities~\eqref{z-gg-intro:ward}. To restore Ward identities we will follow
the regularization procedure outlined in Ref.~\cite{jackiw}, and calculate the
surface term explicitely.

Let
\begin{equation}
 \tilde T^{\mu \nu \lambda}(k, k')
 \to \tilde T^{\mu \nu \lambda}(k, k') + \tilde \Delta^{\mu \nu \lambda}(k, k'),
\end{equation}
where $\tilde \Delta^{\mu \nu \lambda}(k, k')$ is the surface term for the
integral in~\eqref{z-gg:tilde-T}. It is equal to~\cite[(4.21)]{jackiw}
\begin{equation}
 \tilde \Delta^{\mu \nu \lambda}(k, k')
 = \varepsilon^{\mu \nu \lambda \alpha} \tilde w_\alpha,
\end{equation}
where $\tilde w$ is some vector. In the problem under consideration there are
only two linearly independent vectors, $k$ and $k'$, so $\tilde w$ has to be
their linear combination. Let
\begin{equation}
 \tilde w_\alpha = \tilde a(k, k') k_\alpha + \tilde a'(k, k') k'_\alpha,
\end{equation}
where $\tilde a(k, k')$ and $\tilde a'(k, k')$ are some scalars which may depend
on $kk'$, $k^2$ and $k'^2$, and in the general case are not invariant under
interchange $k \leftrightarrow k'$. $T^{\mu \nu \lambda}(k, k')$ is then changed
in the following way:
\begin{equation}
 T^{\mu \nu \lambda}(k, k')
 \to T^{\mu \nu \lambda}(k, k') + \Delta^{\mu \nu \lambda}(k, k'),
 \label{z-gg:regularization}
\end{equation}
where
\begin{equation}
 \begin{split}
  \Delta^{\mu \nu \lambda}(k, k')
  &= \tilde \Delta^{\mu \nu \lambda}(k, k')
   + \tilde \Delta^{\nu \mu \lambda}(k', k)
  \\
  &= \varepsilon^{\mu \nu \lambda \alpha}
     ( (\tilde a (k, k') - \tilde a'(k', k)) k_\alpha
     + (\tilde a'(k, k') - \tilde a (k', k)) k'_\alpha
     )
  \\
  &\equiv
     \varepsilon^{\mu \nu \lambda \alpha}
     (a(k, k') k_\alpha + a'(k, k') k'_\alpha).
 \end{split}
 \label{z-gg:surface-term}
\end{equation}
Note that on shell there is only one Lorentz-invariant scalar depending on
either $k$ or $k'$, namely $kk'$, and in this case $\tilde a(k', k) = \tilde
a(k, k')$, $\tilde a'(k', k) = \tilde a'(k, k')$, hence $a(k, k') = \tilde a(k,
k') - \tilde a'(k, k') = -a'(k, k')$.

Imposing Ward identites~\eqref{z-gg-intro:ward} on $T^{\mu \nu \lambda}(k, k')$
results in the following expressions for $a(k, k')$ and $a'(k, k')$:
\begin{equation}
 a(k, k')
 = -a'(k, k')
 = -\frac{1}{2 \pi^2} \Int_0^1 \Int_0^{1-x}
     \frac{x^2 (1-x) k^2 + z (1-z-xz) k'^2 + 2 x^2 z kk' - (1-x) m^2}
          {2 xz kk' - m^2 + x (1-x) k^2 + z (1-z) k'^2}
    \, \d z \, \d x.
 \label{z-gg:renormalization-a}
\end{equation}
Substitution of \eqref{z-gg:regularization}, \eqref{z-gg:T-unregularized},
\eqref{z-gg:tilde-T-unregularized} into \eqref{z-gg:zgg} provides the
regularized amplitude:
\begin{equation}
 \begin{split}
  \mathcal{M}_{Z \to \gamma \gamma}
  &= \coupling{\psi}{\gamma}^2 \acoupling{\psi}{Z}
     \epsilon_\nu \epsilon'_\mu \tilde \epsilon_\lambda
     T^{\mu \nu \lambda}(k, k')
   \\
  &= \frac{1}{\pi^2} \coupling{\psi}{\gamma}^2 \acoupling{\psi}{Z}
     \epsilon_\nu \epsilon'_\mu \tilde \epsilon_\lambda
     \l(
      \Int_0^1 \Int_0^{1-x}
        \frac{1}{2 xz kk' - m^2 + x (1-x) k^2 + z (1-z) k'^2} \cdot
   \r. \\ & \qquad
       \cdot
        \bigl\{
          [
             \varepsilon^{\nu \lambda \alpha \beta} x ((1-x) k + z k')^\mu
           - \varepsilon^{\mu \lambda \alpha \beta} z (x k + (1-z) k')^\nu
          ] k'_\alpha k_\beta
   \\ & \qquad \l.
        - \varepsilon^{\mu \nu \lambda \alpha}
          \bigl[
           z k' (x k + (1-z) k') k_\alpha - x k ((1-x) k + z k') k'_\alpha
          \bigr]
        \bigr\}
       \, \d z \, \d x
     \r).
 \end{split}
 \label{z-gg:amplitude-off-shell}
\end{equation}
On shell $k^2 = k'^2 = 0$, and
\begin{equation}
 \begin{split}
  \mathcal{M}_{Z \to \gamma \gamma}
  &= \frac{1}{\pi^2} \coupling{\psi}{\gamma}^2 \acoupling{\psi}{Z}
     \epsilon_\nu \epsilon'_\mu \tilde \epsilon_\lambda
     \l(
        \l(
           k^\mu  \varepsilon^{\nu \lambda \alpha \beta}
         - k'^\nu \varepsilon^{\mu \lambda \alpha \beta}
        \r) k'_\alpha k_\beta
        \Int_0^1 \Int_0^1 \frac{x (1-x) \, \d z \, \d x}{2 xz kk' - m^2}
  \r. \\ & \qquad \l.
      + \Bigl(
           \l(
              k'^\mu \varepsilon^{\nu \lambda \alpha \beta}
            - k^\nu  \varepsilon^{\mu \lambda \alpha \beta}
           \r) k'_\alpha k_\beta
         - \varepsilon^{\mu \nu \lambda \alpha} kk' (k - k')_\alpha
        \Bigr)
        \Int_0^1 \Int_0^{1-x} \frac{xz \, \d z \, \d x}{2 xz kk' - m^2}
     \r).
 \end{split}
\end{equation}
Comparing this expression with~\eqref{z-gg-intro:T-onshell} we obtain:
\begin{align}
 N - M'
 &= -\Int_0^1 \Int_0^{1-x} \frac{xz \, \d x \, \d z}{2 xz kk' - m^2}
  = -\frac{1}{2}
    \l(
     1 + \eta_Z \ln^2 \l( 1 - \frac{1 + \sqrt{1 - 4 \eta_Z}}{2 \eta_Z} \r)
    \r),
 \\
 N' - M
 &= -\Int_0^1 \Int_0^{1-x} \frac{x (1-x) \, \d x \, \d z}{2 xz kk' - m^2}
  = \frac{\sqrt{1 - 4 \eta_Z}}{2}
    \ln \l( \frac{1 + \sqrt{1 - 4 \eta_Z}}{2 \eta_Z} - 1 \r) - 1,
\end{align}
where
\begin{equation}
 \eta_Z = \l( \frac{m}{m_Z} \r)^2,
\end{equation}
$m$ is the fermion mass, $m_Z$ is the Z boson mass.

\section{Axial anomaly in $Z \to \gamma \gamma$ transition amplitude}

\subsection{The Glashow-Weinberg-Salam theory of weak interactions with massless
fermions}

\label{gws}

Let us briefly remind the mechanism of cancellation of anomalies in the
Glashow-Weinberg-Salam theory of weak interactions~\cite{bouchiat}.
We will consider one generation of fermions in unbroken $SU(2) \times U(1)$
theory leaving consideration of spontaneously broken theory to the next
subsection.  The lagrangian of the theory is
\begin{equation}
 \begin{split}
   \L
   &= \row{\bar \nu}{\bar e} i \s D_l \frac{1 + \gamma^5}{2} \column{\nu}{\e}
    + \bar e i \s D_e \frac{1 - \gamma^5}{2} e
   \\
   &+ \row{\bar u}{\bar d} i \s D_q \frac{1 + \gamma^5}{2} \column{u}{d}
    + \bar u i \s D_u \frac{1 - \gamma^5}{2} u
    + \bar d i \s D_d \frac{1 - \gamma^5}{2} d
   \\
   &- \frac{1}{4} F_a^{\mu \nu} F_{\mu \nu}^a
    - \frac{1}{4} F_{\mu \nu}   F^{\mu \nu},
 \end{split}
 \label{gws:lagrangian}
\end{equation}
where
\begin{gather}
 \begin{aligned}
  D_{\mu L} &= \p_\mu - i g A_\mu^a \tau^a - i g' B_\mu Y_L, && L \in \{l, q\},
  \\
  D_{\mu R} &= \p_\mu - i g' B_\mu Y_R, && R \in \{e, u, d\},
 \end{aligned}
 \notag
 \\
        Y_l = -\frac{1}{2},
  \quad Y_e = -1,
  \quad Y_q = \frac{1}{6},
  \quad Y_u = \frac{2}{3},
  \quad Y_d = -\frac{1}{3},
  \label{gws:hypercharges}
 \\
       F^a_{\mu \nu} = \p_\mu A_\nu^a - \p_\nu A_\mu^a
                     + g \varepsilon^{abc} A_\mu^b A_\nu^c,
 \quad F_{\mu \nu}   = \p_\mu B_\nu   - \p_\nu B_\mu.
 \notag
\end{gather}
It is invariant under $SU(2) \times U(1)$ transformations:
\begin{align}
 SU(2) \colon
 && A_\mu^a(x) & \to A_\mu^a(x)
                     + \tfrac{1}{g} \p_\mu \alpha^a(x)
                     + \varepsilon^{abc} A_\mu^b(x) \alpha^c(x),
               & \psi_L(x)  & \to \e^{-i g \tau^a \alpha^a(x)} \psi_L(x),
               & L          & \in \{l, q\},
 \label{gws:su2}
 \\
 U(1)  \colon
 && B_\mu(x)   & \to B_\mu(x) + \tfrac{1}{g'} \p_\mu \beta(x),
               & \psi_X(x)  & \to \e^{-i g' Y_X \beta(x)} \psi_X(x),
               & X          & \in \{l, q, e, u, d\},
 \label{gws:u1}
\end{align}
where
\begin{gather*}
 \psi_l = \frac{1 + \gamma^5}{2} \column{\nu}{e},
 \quad
 \psi_q = \frac{1 + \gamma^5}{2} \column{u}{d},
 \\
 \psi_R = \frac{1 - \gamma^5}{2} R, \quad R \in \{ e, u, d \}.
\end{gather*}
These transformations give rise to the following N\"other currents:
\begin{align}
 SU(2) \colon && j_\lambda^a
 &= \row{\bar \nu}{\bar e} \gamma_\lambda \tau^a
    \tfrac{1 + \gamma^5}{2} \column{\nu}{e}
  + \row{\bar u}{\bar d}   \gamma_\lambda \tau^a
    \tfrac{1 + \gamma^5}{2} \column{u}{d},
 \label{gws:ja}
 \\
 U(1) \colon && j_\lambda
 &= \row{\bar \nu}{\bar e} \gamma_\lambda Y_l
    \tfrac{1 + \gamma^5}{2} \column{\nu}{e}
  + \bar e \gamma_\lambda Y_e \tfrac{1 - \gamma^5}{2} e
  + \row{\bar u}{\bar d} \gamma_\lambda Y_q
    \tfrac{1 + \gamma^5}{2} \column{u}{d}
  + \bar u \gamma_\lambda Y_u \tfrac{1 - \gamma^5}{2} u
  + \bar d \gamma_\lambda Y_d \tfrac{1 - \gamma^5}{2} d,
 \label{gws:j}
\end{align}
where terms depending on boson fields were omitted.

The switch from bare gauge bosons to states with definite masses is performed
with the help of the following relations:
\begin{equation}
 W_\mu^\pm = \frac{A_\mu^1 \mp i A_\mu^2}{\sqrt{2}},
 \quad
 Z_\mu = A_\mu^3 \cos \theta - B_\mu \sin \theta,
 \quad
 A_\mu = A_\mu^3 \sin \theta + B_\mu \cos \theta,
\end{equation}
where
\begin{equation}
 \cos \theta = \frac{g}{\sqrt{g^2 + g'^2}},
 \quad
 \sin \theta = \frac{g'}{\sqrt{g^2 + g'^2}},
\end{equation}
$\theta$ is the Weinberg angle or electroweak mixing angle. The current coupled
to Z~boson is
\begin{equation}
 j_\lambda^Z = j_\lambda^3 \cos^2 \theta - j_\lambda \sin^2 \theta,
\end{equation}
or, explicitely,
\begin{multline}
 j_\lambda^Z
 = \tfrac{1}{2} \bar \nu \gamma_\lambda \tfrac{1 + \gamma^5}{2} \nu
 - \l( \tfrac{1}{4} \cos^2 \theta - \tfrac{3}{4} \sin^2 \theta \r)
   \bar e \gamma_\lambda e
 - \tfrac{1}{4} \bar e \gamma_\lambda \gamma^5 e
 + \l( \tfrac{3}{12} \cos^2 \theta - \tfrac{5}{12} \sin^2 \theta \r)
   \bar u \gamma_\lambda u
 + \tfrac{1}{4} u \gamma_\lambda \gamma^5 u
 \\
 - \l( \tfrac{3}{12} \cos^2 \theta - \tfrac{1}{12} \sin^2 \theta \r)
   \bar d \gamma_\lambda d
 - \tfrac{1}{4} \bar d \gamma_\lambda \gamma^5 d.
 \label{gws:jZ}
\end{multline}

According to the N\"other's theorem, currents~\eqref{gws:ja},~\eqref{gws:j} have
to be conserved, therefore, $j_\mu^Z$ has to be conserved as well:
\begin{equation}
 \p^\lambda j_\lambda^Z = 0.
 \label{gws:djZ}
\end{equation}
Let us check this statement with explicit calculation. Consider the matrix
element $\braket{A(k), A(k')|j_\lambda^Z(x)|0}$. For each term in~\eqref{gws:jZ}
there is a term in the leading-order approximation of this element represented
by a sum of diagrams~\ref{diag:z-gg} with the corresponding fermion $\psi$
running through the loops. Terms with vector coupling between $Z$ and $\psi$
give zero according to Furry's theorem. Terms with axial coupling have the
following structure:
\begin{equation}
 \int
  \e^{-iqx} \braket{A(k), A(k')| \bar \psi \gamma^\lambda \gamma^5 \psi |0}
 \, \d^4 x
 = (2 \pi)^4 \delta^{(4)}(k + k' - q)
   \vcoupling{\psi}{A}^2
   \epsilon_\mu \epsilon'_\nu T_\psi^{\mu \nu \lambda}(k, k'),
\end{equation}
where square brackets denote coupling constants between the fermion and bosons;
index $V$ designates vector coupling, $A$ will be used for axial. $T_\psi^{\mu
\nu \lambda}(k, k')$ is the $T^{\mu \nu \lambda}(k, k')$ tensor defined
by~\eqref{z-gg:amplitude-off-shell} indexed with $\psi$ to indicate its
dependence on the type of fermion in the loop. However, in the theory under
consideration there is no such dependence, since all fermions are massless. The
derivative of the matrix element is:
\begin{equation}
 \tfrac{g}{\cos \theta}
 \int
  \e^{-iqx} \braket{A(k), A(k')| \p^\lambda j_\lambda^Z |0}
 \, \d^4 x
 =     (2 \pi)^4 \delta^{(4)}(k + k' - q)
 \cdot \mathcal{A}_Z(A(k), A(k')),
\end{equation}
where the factor $\tfrac{g}{\cos \theta}$ is extracted for convenience, and
\begin{equation}
 \mathcal{A}_Z(A(k), A(k'))
 = \epsilon_\mu \epsilon'_\nu q_\lambda T^{\mu \nu \lambda}(k, k')
   \Sum_\psi \acoupling{\psi}{Z} \vcoupling{\psi}{A}^2
 \label{gws:anomaly}
\end{equation}
is an anomaly. It should be zero to be consistent with~\eqref{gws:djZ}. Let us
check the third Ward identity~\eqref{z-gg-intro:ward3}.  Replacing $\tilde
\epsilon_\lambda$ by $q_\lambda$ in~\eqref{z-gg:amplitude-off-shell} and noticing
the similarity between the resulting expression
and~\eqref{phi-gg:amplitude-off-shell}, we obtain:
\begin{equation}
 q_\lambda T^{\mu \nu \lambda}(k, k')
 = - \frac{1}{2 \pi^2} \varepsilon^{\mu \nu \alpha \beta} k'_\alpha k_\beta
   + 2 m T^{\mu \nu}(k, k').
 \label{gws:ward3}
\end{equation}
As was mentioned above, $m = 0$, so
\begin{equation}
 q_\lambda T^{\mu \nu \lambda}(k, k')
 = - \frac{1}{2 \pi^2} \varepsilon^{\mu \nu \alpha \beta} k'_\alpha k_\beta
 \ne 0.
\end{equation}
Consequently, in order for the anomaly to cancel out, the sum over fermions
in~\eqref{gws:anomaly} has to be zero. Using coupling constants presented in
Table~\ref{gws:couplings}, we get:
\begin{equation}
 \Sum_\psi \acoupling{\psi}{Z} \vcoupling{\psi}{A}^2
 = \tfrac{g^3 \sin^2 \theta}{4 \cos \theta}
 \cdot \l(
        - 1 \cdot 0^2
        + 1 \cdot (-1)^2
        + 3 \cdot \l( 
                     - 1 \cdot \l(  \tfrac{2}{3} \r)^2
                     + 1 \cdot \l( -\tfrac{1}{3} \r)^2
                  \r)
       \r)
  = 0,
\end{equation}
where the multiplier 3 takes into account the three colors of quarks.

In a similar way one can prove cancellation of anomalies for other N\"other
currents as well.

\begin{table}[h]
 \centering
 \renewcommand{\arraystretch}{1.6}
 \begin{tabular}{|c||c|c|c|c|} \hline
  $\psi$ & $\nu$ & $e$ & $u$ & $d$ 
  \\ \hline
    $\vcoupling{\psi}{A}$
  & 0
  & $-g \sin \theta$
  & $\tfrac{2}{3} g \sin \theta$
  & $-\tfrac{1}{3} g \sin \theta$
  \\ \hline
    $\acoupling{\psi}{Z}$
  & $-\tfrac{g}{4 \cos \theta}$
  & $ \tfrac{g}{4 \cos \theta}$
  & $-\tfrac{g}{4 \cos \theta}$
  & $ \tfrac{g}{4 \cos \theta}$
  \\ \hline
 \end{tabular}
 \caption{Coupling constants appearing in eq.~\eqref{gws:anomaly}.}
 \label{gws:couplings}
\end{table}

\subsection{The Glashow-Weinberg-Salam theory with massive fermions}

\label{mgws}

The fact that fermions are massless was crucial in the cancellation of the
anomaly presented in the previous section.
Let us consider now what happens when we take into account spontaneous symmetry breaking and
introduce the Higgs mechanism of fermions and gauge bosons mass generation. We
keep working with a single generation of fermions, but the arguments will be
independent of quark mixing, and can be readily generalized to a theory with any
number of generations. We also will consider only the neutral current coupled to
Z~boson, but the same reasoning should be applicable to charged currents as
well.

The Standard Model lagrangian is
\begin{equation}
 \begin{split}
  \L
  &= \abs{D_\mu H}^2
   - \tfrac{\lambda^2}{2} \l( H^\dagger H - \tfrac{v^2}{2} \r)^2
   - \tfrac{1}{4} F_a^{\mu \nu} F_{\mu \nu}^a
   - \tfrac{1}{4} F^{\mu \nu} F_{\mu \nu} 
  \\
  &+ \row{\bar \nu}{\bar e} i \s D_l
     \tfrac{1 + \gamma^5}{2} \column{\bar \nu}{\bar e}
   + \bar e i \s D_e \tfrac{1 - \gamma^5}{2} e
  \\
  &+ \row{\bar u}{\bar d} i \s D_q
     \tfrac{1 + \gamma^5}{2} \column{u}{d}
   + \bar u i \s D_u \tfrac{1 - \gamma^5}{2} u
   + \bar d i \s D_d \tfrac{1 - \gamma^5}{2} d
  \\
  &- \lambda_e \row{\bar \nu}{\bar e} H \tfrac{1 - \gamma^5}{2} e
   - \lambda_e \bar e H^\dagger \tfrac{1 + \gamma^5}{2} \column{\nu}{e}
  \\
  &- \lambda_u \row{-\bar d}{\bar u} H^* \tfrac{1 - \gamma^5}{2} u
   - \lambda_u \bar u H^T \tfrac{1 + \gamma^5}{2} \column{-d}{u}
  \\
  &- \lambda_d \row{\bar u}{\bar d} H \tfrac{1 - \gamma^5}{2} d
   - \lambda_d \bar d H^\dagger \tfrac{1 + \gamma^5}{2} \column{u}{d},
  \\
 \end{split}
 \label{mgws:lagrangian}
\end{equation}
where
\begin{gather}
 H(x) = \frac{1}{\sqrt{2}} \column{\phi(x)}{v + h(x) + i \chi(x)},
 \\
 D_\mu = \p_\mu - i g A_\mu^a \tau^a - i g' B_\mu Y, \quad Y = \tfrac{1}{2}.
 \notag
\end{gather}
This lagrangian is invariant under $SU(2) \times U(1)$
transformations~\eqref{gws:su2}, \eqref{gws:u1} with the following
transformations for the Higgs field:
\begin{align}
 SU(2) \colon & H(x) \to \e^{-i g \tau^a \alpha^a(x)} H(x), \\
 U(1)  \colon & H(x) \to \e^{-i g' Y \beta(x)} H(x).
\end{align}
Gauge currents~\eqref{gws:ja} and~\eqref{gws:j} get additional terms which
depend on $H$:
\begin{align}
 SU(2) \colon && j_\mu^a
 &=
    \row{\bar \nu}{\bar e} \gamma_\mu \tau^a
    \tfrac{1 + \gamma^5}{2} \column{\nu}{e}
  + \row{\bar u}{\bar d}   \gamma_\mu \tau^a
    \tfrac{1 + \gamma^5}{2} \column{u}{d}
  + i (H^\dagger \tau^a D_\mu H - (D_\mu H)^\dagger \tau^a H),
 \\
 U(1) \colon && j_\mu
 &=
    \row{\bar \nu}{\bar e} \gamma_\mu Y_l
    \tfrac{1 + \gamma^5}{2} \column{\nu}{e}
  + \bar e \gamma_\mu Y_e \tfrac{1 - \gamma^5}{2} e
  + \row{\bar u}{\bar d} \gamma_\mu Y_q
    \tfrac{1 + \gamma^5}{2} \column{u}{d}
  + \bar u \gamma_\mu Y_u \tfrac{1 - \gamma^5}{2} u
  + \bar d \gamma_\mu Y_d \tfrac{1 - \gamma^5}{2} d
 \notag \\ &&& \quad
  + \, i (H^\dagger Y D_\mu H - (D_\mu H)^\dagger Y H).
\end{align}

Explicitely, covariant derivatives have the following representation:
\begin{align}
 D_{\mu L} &= \begin{pmatrix}
    \p_\mu
  - \frac{ig \sin \theta}{2} (1 + 2 Y_L) A_\mu
  - \frac{ig (\cos^2 \theta - 2 Y_L \sin^2 \theta)}{2 \cos \theta} Z_\mu
  & -\frac{ig}{\sqrt{2}} W_\mu^+
  \\
    -\frac{ig}{\sqrt{2}} W_\mu^-
  & \p_\mu
  + \frac{ig}{2} \sin \theta (1 - 2 Y_L) A_\mu
  + \frac{ig (\cos^2 \theta + 2 Y_L \sin^2 \theta)}{2 \cos \theta} Z_\mu
 \end{pmatrix},
 \label{mgws:derivative-l}
 \\
 D_{\mu R}
 &= \p_\mu
 - ig Y_R \sin \theta A_\mu
 + \frac{ig Y_R \sin^2 \theta}{\cos \theta} Z_\mu
 \label{mgws:derivative-r}
\end{align}
In the case of the Higgs field $Y = \tfrac{1}{2}$, and
\begin{equation}
 D_\mu = \begin{pmatrix}
  \p_\mu - i g \sin \theta A_\mu - \tfrac{ig \cos 2 \theta}{2 \cos \theta} Z_\mu
  &
  -\tfrac{ig}{\sqrt{2}} W_\mu^+
  \\
  -\tfrac{ig}{\sqrt{2}} W_\mu^-
  &
  \p_\mu + \tfrac{ig}{2 \cos \theta} Z_\mu
 \end{pmatrix}.
 \label{mgws:derivative-h}
\end{equation}
Consequently, the Higgs field introduces the following additional terms in the
current~\eqref{gws:jZ}:
\begin{multline}
 j_\lambda^Z
 = \tfrac{1}{2} \bar \nu \gamma_\lambda \tfrac{1 + \gamma^5}{2} \nu
 - \l( \tfrac{1}{4} \cos^2 \theta - \tfrac{3}{4} \sin^2 \theta \r)
   \bar e \gamma_\lambda e
 - \tfrac{1}{4} \bar e \gamma_\lambda \gamma^5 e
 + \l( \tfrac{3}{12} \cos^2 \theta - \tfrac{5}{12} \sin^2 \theta \r)
   \bar u \gamma_\lambda u
 + \tfrac{1}{4} u \gamma_\lambda \gamma^5 u
 \\
 - \l( \tfrac{3}{12} \cos^2 \theta - \tfrac{1}{12} \sin^2 \theta \r)
   \bar d \gamma_\lambda d
 - \tfrac{1}{4} \bar d \gamma_\lambda \gamma^5 d
 + \tfrac{i}{4} \cos 2 \theta (\phi^* \p_\lambda \phi - \phi \p_\lambda \phi^*)
 + \tfrac{g \sin \theta \cos 2 \theta}{2} A_\lambda \phi^* \phi
 \\
 + \tfrac{g \cos^2 2 \theta}{4 \cos \theta} Z_\lambda \phi^* \phi
 - \tfrac{ g \sin^2 \theta}{2 \sqrt{2}}
   (v + h) (W_\lambda^+ \phi^* + W_\lambda^- \phi)
 + \tfrac{ig \sin^2 \theta}{2 \sqrt{2}} \chi
   (W_\lambda^+ \phi^* - W_\lambda^- \phi)
 \\
 + \tfrac{1}{2} (v + h) \p_\lambda \chi
 - \tfrac{1}{2} \chi \p_\lambda h
 + \tfrac{g}{4 \cos \theta} Z_\lambda \bigl( (v + h)^2 + \chi^2 \bigr).
 \label{mgws:jz}
\end{multline}
The first term in the last line, $\tfrac{1}{2} (v + h) \p_\lambda \chi$, indicates
a term in the lagrangian which mixes Z~boson and the Golstone boson~$\chi$:
$\tfrac{v}{2} Z_\lambda \p^\lambda \chi$. The corresponding vertex is
represented in fig.~\ref{diag:z-chi}.
\begin{figure}[h]
 \centering
 \begin{tikzpicture}[>=latex]
  \draw[boson]  (0, 0) -- (1, 0);
  \draw[scalar] (1, 0) -- (2, 0) node [right] {
   $ = i \bcoupling{\mathrm{Z}}{\chi} \cdot (-i (-q_\lambda))
     = - \bcoupling{\mathrm{Z}}{\chi} q_\lambda$
  };
  \draw[->,auto] (0.25, 2mm) to node {$q$} (0.75, 2mm);
 \end{tikzpicture}
 \caption{Vertex $Z_\lambda \p^\lambda \chi$.}
 \label{diag:z-chi}
\end{figure}
It results in appearance of two extra anomalous diagrams~\ref{diag:z-chi-gg} in
addition to those shown in fig.~\ref{diag:z-gg}.
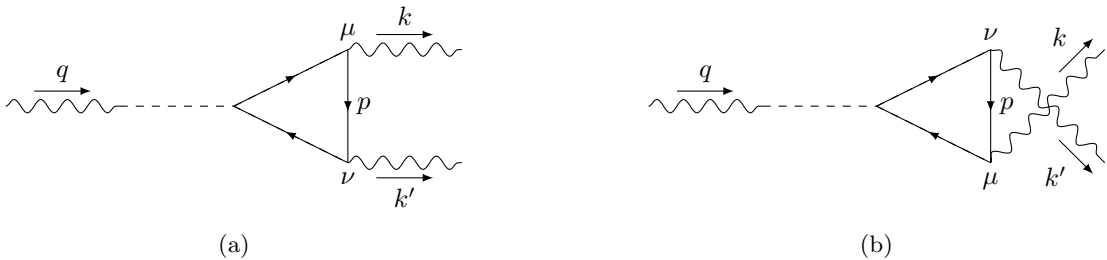
\begin{figure}[h]
 \centering
 \begin{subfigure}{0.4\textwidth}
  \centering
   \begin{tikzpicture}[>=latex]
    \begin{scope}[shape=coordinate]
     \node (q) {}
      [grow=right]
      child {
       node (h) {}
       child {
        node (lambda) {}
        child { node [label={[shape=rectangle]below:$\nu$}] (nu) {}
                child {node (k') {} edge from parent [boson]}}
        child { node [label={[shape=rectangle]above:$\mu$}] (mu) {}
                child {node (k)  {} edge from parent [boson]}}
        edge from parent [scalar]
       }
       edge from parent [boson]
      };
    \end{scope}

    \draw[fermion,auto] (lambda) -- (mu);
    \draw[fermion,auto] (mu) to node {$p$} (nu);
    \draw[fermion,auto] (nu) -- (lambda);

    \draw[->,auto] ([yshift=2mm]$(q)!0.25!(h)$)
                   to node {$q$}
                   ([yshift=2mm]$(q)!0.75!(h)$);
    \draw[->,auto] ([yshift=2mm]$(mu)!0.25!(k)$)
                   to node {$k$}
                   ([yshift=2mm]$(mu)!0.75!(k)$);
    \draw[->,auto] ([yshift=-2mm]$(nu)!0.25!(k')$)
                   to node [swap] {$k'$}
                   ([yshift=-2mm]$(nu)!0.75!(k')$);

    \coordinate (diagramne) at (current bounding box.north east);
    \coordinate (diagramsw) at (current bounding box.south west);
   \end{tikzpicture}

  \caption{ }
  \label{diag:z-chi-gg-a}
 \end{subfigure}
 ~
 \begin{subfigure}{0.5\textwidth}
  \centering

  \begin{tikzpicture}[>=latex]
   \useasboundingbox (diagramne) rectangle (diagramsw);

   \begin{scope}[shape=coordinate]
    \node (q) {}
     [grow=right]
     child {
      node (h) {}
      child {
       node (lambda) {}
       child {
        node [label={[shape=rectangle]below:$\mu$}] (mu) {}
        child {node (k')  {} edge from parent [draw=none]}
       }
       child {
        node [label={[shape=rectangle]above:$\nu$}] (nu) {}
        child {node (k) {} edge from parent [draw=none]}
       }
       edge from parent [scalar]
      }
      edge from parent [boson]
     };
   \end{scope}

   \draw[boson] (mu) -- (k);
   \draw[boson] (nu) -- (k');

   \draw[fermion,auto] (lambda) -- (nu);
   \draw[fermion,auto] (nu) to node {$p$} (mu);
   \draw[fermion,auto] (mu) -- (lambda);

   \draw[->,auto] ([yshift=2mm]$(q)!0.25!(h)$)
                  to node {$q$}
                  ([yshift=2mm]$(q)!0.75!(h)$);
   \draw[->,auto] ([yshift=3mm]$(mu)!0.6!(k)$)
                  to node {$k$}
                  ([yshift=3mm]$(mu)!0.9!(k)$);
   \draw[->,auto] ([yshift=-3mm]$(nu)!0.6!(k')$)
                  to node [swap] {$k'$}
                  ([yshift=-3mm]$(nu)!0.9!(k')$);
  \end{tikzpicture}

  \caption{ }
  \label{diag:z-chi-gg-b}
 \end{subfigure}
 \caption{Additional lowest-order Feynman diagrams of $\mathrm{Z} \to \gamma
 \gamma$ transition appearing due to vertex in fig.~\ref{diag:z-chi}.}
 \label{diag:z-chi-gg}
\end{figure}
Anomaly of the derivative of the matrix element $\braket{A(k), A(k')|
j_\lambda^Z |0}$ in this case is
\begin{equation}
 \mathcal{A}_Z(A(k), A(k'))
 = \epsilon_\mu \epsilon'_\nu q_\lambda
   \Sum_\psi \l\{
      \acoupling{\psi}{Z} \vcoupling{\psi}{A}^2 T_\psi^{\mu \nu \lambda}(k, k')
    +       (- \bcoupling{Z}{\chi} q^\lambda)
      \cdot \frac{i}{q^2}
      \cdot \acoupling{\psi}{\chi} \vcoupling{\psi}{A}^2 T_\psi^{\mu \nu}(k, k')
   \r\},
\end{equation}
where the first term is a sum of derivatives of diagrams in
fig.~\ref{diag:z-gg}, \eqref{gws:anomaly}, and the second term is a sum of
derivatives of diagrams in fig.~\ref{diag:z-chi-gg}, expressed as a derivative
of a product of vertex represented in fig.~\ref{diag:z-chi}, a propagator of
$\chi$, and a sum of diagrams in fig.~\ref{diag:phi-gg}.
$T_\psi^{\mu \nu \lambda}(k, k')$ and $T_\psi^{\mu \nu}(k, k')$ are defined
by~\eqref{z-gg:amplitude-off-shell} and~\eqref{phi-gg:amplitude-off-shell}, and
do depend on fermion mass. With the help of~\eqref{gws:ward3}, this expression
can be rewritten as follows:
\begin{multline}
 \mathcal{A}_Z(A(k), A(k'))
 = - \frac{1}{2 \pi^2} \varepsilon^{\mu \nu \alpha \beta}
     \epsilon_\mu \epsilon'_\nu k'_\alpha k_\beta
     \Sum_\psi \acoupling{\psi}{Z} \vcoupling{\psi}{A}^2
 \\
   + \epsilon_\mu \epsilon'_\nu
     \Sum_\psi \vcoupling{\psi}{A}^2
      \Bigl(
         2 m_\psi \acoupling{\psi}{Z}
       - i \bcoupling{Z}{\chi} \acoupling{\psi}{\chi}
      \Bigr) T_\psi^{\mu \nu}(k, k').
 \label{mgws:anomaly}
\end{multline}
According to the previous section, the first sum is zero. Let us consider the
second sum. Axial coupling constants between $\chi$ and fermions, as well as
fermion masses, follow from the last terms of the
lagrangian~\eqref{mgws:lagrangian}. Keeping only the terms with $v$ or $\chi$,
we obtain:
\begin{equation}
 \begin{split}
   - \lambda_e   \row{\bar \nu}{\bar e} H \frac{1 - \gamma^5}{2} e
   - \lambda_e \bar e H^\dagger \frac{1 + \gamma^5}{2} \column{\nu}{e}
  &= \ldots
   - \lambda_e \bar e \frac{v + i \chi}{\sqrt{2}} \frac{1 - \gamma^5}{2} e
   - \lambda_e \bar e \frac{v - i \chi}{\sqrt{2}} \frac{1 + \gamma^5}{2} e
  \\
  &= \ldots - m_e \bar e e - \frac{i m_e}{v} \chi \bar e \gamma^5 e,
 \end{split}
\end{equation}
where
\begin{equation}
 m_e = \frac{\lambda_e v}{\sqrt{2}}.
\end{equation}
By analogy,
\begin{align}
 - \lambda_d \row{\bar u}{\bar d} H \frac{1 - \gamma^5}{2} d
 - \lambda_d \bar d H^\dagger \frac{1 + \gamma^5}{2} \column{u}{d}
 &= \ldots - m_d \bar d d - \frac{i m_d}{v} \chi \bar d \gamma^5 d,
 & m_d &= \frac{\lambda_d v}{\sqrt{2}},
 \\
 - \lambda_u \row{-\bar d}{\bar u} H^* \frac{1 - \gamma^5}{2} u
 - \lambda_u \bar u H^T \frac{1 + \gamma^5}{2} \column{-d}{u}
 &= \ldots - m_u \bar u u + \frac{i m_u}{v} \chi \bar u \gamma^5 u,
 & m_u &= \frac{\lambda_u v}{\sqrt{2}}.
\end{align}
Coupling constants are summarized in Table~\ref{mgws:couplings}. With its help we
find out that the following equation holds true:
\begin{equation}
 2 m_\psi \acoupling{\psi}{Z} - i \bcoupling{Z}{\chi} \acoupling{\psi}{\chi} = 0,
 \quad \psi \in \{ \nu, e, u, d \}.
\end{equation}
Consequently, \eqref{mgws:anomaly} is equal to zero, and the Z~boson current is
conserved.
\begin{table}[h]
 \centering
 \renewcommand{\arraystretch}{1.6}
 \begin{tabular}{|c||c|c|c|c|}
  \hline
  $\psi$ & $\nu$ & $e$ & $u$ & $d$
  \\ \hline
    $\acoupling{\psi}{Z}$
  & $-\tfrac{g}{4 \cos \theta}$ & $\tfrac{g}{4 \cos \theta}$
  & $-\tfrac{g}{4 \cos \theta}$ & $\tfrac{g}{4 \cos \theta}$
  \\ \hline
    $\acoupling{\psi}{\chi}$
  & 0 & $-\tfrac{i m_e}{v}$ & $\tfrac{i m_u}{v}$ & $-\tfrac{i m_d}{v}$
  \\ \hline
 \end{tabular}
 \\[1ex]
 $\bcoupling{Z}{\chi} = \frac{g v}{2 \cos \theta}$
 \caption{Coupling constants appearing in eq.~\eqref{mgws:anomaly}.}
 \label{mgws:couplings}
\end{table}

\section{Conclusions}

\zgg{} decay is once again proven to be impossible due to properties of
polarization vectors of Z~boson and photons and Bose statistics.  Of course,
it does not necessarily mean that we should stop looking for it. On the
contrary, discovery of such a decay would provide evidence for physics beyond
our current understanding of the quantum field theory. However, the hope is
quite small, and we probably should not waste too much resources on it. In
this aspect its investigation is very similar to the search of
faster-than-light neutrinos or violations of CPT theorem.

To be renormalizable, a theory has to be free of axial anomalies in gauge
(N\"other) currents. The mechanism of cancellation of anomalies in the
Standard Model has been explicitely demonstrated. Its key features are:
\begin{itemize}
 \item Anomaly gives two terms in an amplitude, one depends on fermion masses,
  and the other does not.
 \item Mass-independent term vanishes in the same way as in the theory with
  massless fermions (hypercharge values are such that the corresponding sum of
  coupling constants vanishes).
 \item Spontaneous symmetry breaking not only generates fermion masses, but
  also mixes Z~boson with Goldstone boson. Thus fermion masses and Higgs boson
  vacuum expectation value appear in Goldstone boson coupling constants.
 \item Goldstone boson provides an additional term in the amplitude which
  exactly cancel out the mass-dependent term of the anomaly.
\end{itemize}

\section{Acknowledgements}

I would like to express my deep gratitude to Prof.~M.~I.~Vysotsky for his
willingness to give his time to introduce me into the field of particle physics
and patient guidance during the course of this work. I am partially supported by
RFFI grants 12-02-00193-a and 14-02-00995.

\appendix

\section{Trace of six Dirac matrices times $\gamma^5$}

An expression for $\tr \gamma^a \gamma^b \gamma^c \gamma^d \gamma^e
\gamma^f \gamma^5$ can be easily derived. In four dimensions there are four
different Dirac matrices, however in this expression there are six factors.
Therefore, some factors have to be the same. There cannot be only two equal
factors since it would require five different matrices.  Let there be two pairs
of equal factors, for instance, let $a = b$ and $c = d$. Then
\begin{equation}
 \tr \gamma^a \gamma^b \gamma^c \gamma^d \gamma^e \gamma^f \gamma^5
 = \tfrac{1}{4} \tr \bigl(
    \{ \gamma^a, \gamma^b \} \{ \gamma^c, \gamma^d \} \gamma^e \gamma^f \gamma^5
   \bigr)
 = g^{ab} g^{cd} \tr \gamma^e \gamma^f \gamma^5
 = 0.
\end{equation}
Choice of other pairs of equal factors results only in change of the sign of the
resulting expression due to swapping of gamma matrices, so the trace keeps being
equal to zero. Obviously, if there are three pairs of equal factors, the result
equals zero again.

Consider the case of three equal factors. First, let $d$, $e$ and $f$ be
different.  Then among $a$, $b$ and $c$ there are two or three equal indices.
Let $a = b \ne c$. Then
\begin{equation}
 \tr \gamma^a \gamma^b \gamma^c \gamma^d \gamma^e \gamma^f \gamma^5
 = g^{ab} \tr \gamma^c \gamma^d \gamma^e \gamma^f \gamma^5
 = 4i g^{ab} \varepsilon^{cdef}.
 \label{tr7:ab}
\end{equation}
Now, let $a = c \ne b$. Then the right hand side of~\eqref{tr7:ab} is equal to
zero. However, the following expression is valid in both cases:
\begin{equation}
 \tr \gamma^a \gamma^b \gamma^c \gamma^d \gamma^e \gamma^f \gamma^5
 = 4i \l( g^{ab} \varepsilon^{cdef} - g^{ac} \varepsilon^{bdef} \r).
 \label{tr7:ab+ac}
\end{equation}
Let $b = c \ne a$. Then, once again, right hand sides of both~\eqref{tr7:ab} and
~\eqref{tr7:ab+ac} equals zero, but the following equation is valid in all three
cases:
\begin{equation}
 \tr \gamma^a \gamma^b \gamma^c \gamma^d \gamma^e \gamma^f \gamma^5
 = 4i \l(  
        g^{ab} \varepsilon^{cdef}
      - g^{ac} \varepsilon^{bdef}
      + g^{bc} \varepsilon^{adef}
   \r).
 \label{tr7:ab+ac+bc}
\end{equation}
It is also valid when $a = b = c$.

Finally, let $a$, $b$ and $c$ be different. Applying the same reasoning to $d$,
$e$ and $f$, and noting that the right hand side of~\eqref{tr7:ab+ac+bc} equals
zero, we get the final result
\begin{equation}
 \tr \gamma^a \gamma^b \gamma^c \gamma^d \gamma^e \gamma^f \gamma^5
 = 4i \l(  
        g^{ab} \varepsilon^{cdef}
      - g^{ac} \varepsilon^{bdef}
      + g^{bc} \varepsilon^{adef}
      + g^{de} \varepsilon^{abcf}
      - g^{df} \varepsilon^{abce}
      + g^{ef} \varepsilon^{abcd}
   \r),
 \label{tr7:tr7}
\end{equation}
which is valid for any combination of $\gamma$ matrices.

Moving the $\gamma^a$ matrix through the trace, a useful identity can be
obtained:
\begin{equation}
 \begin{split}
   \tr \gamma^a \gamma^b \gamma^c \gamma^d \gamma^e \gamma^f \gamma^5
  &= 2 g^{ab} \tr \gamma^c \gamma^d \gamma^e \gamma^f \gamma^5
   - \tr \gamma^b \gamma^a \gamma^c \gamma^d \gamma^e \gamma^f \gamma^5
  \\
  &= \ldots
  \\
  &= 8i \l(
          g^{ab} \varepsilon^{cdef}
        - g^{ac} \varepsilon^{bdef}
        + g^{ad} \varepsilon^{bcef}
        - g^{ae} \varepsilon^{bcdf}
        + g^{af} \varepsilon^{bcde}
     \r)
  \\
  &  \quad + \tr \gamma^b \gamma^c \gamma^d \gamma^e \gamma^f \gamma^5 \gamma^a.
 \end{split}
\end{equation}
Since the last trace is equal to the first one, the expression in the
parentheses equals zero:
\begin{equation}
   g^{ab} \varepsilon^{cdef}
 - g^{ac} \varepsilon^{bdef}
 + g^{ad} \varepsilon^{bcef}
 - g^{ae} \varepsilon^{bcdf}
 + g^{af} \varepsilon^{bcde}
 = 0.
 \label{tr7:identity}
\end{equation}


\begin{thebibliography}{9}
 \bibitem{landau}
  L.D. Landau.
  {\it On total angular momentum of a system of two photons.}
  Doklad Akademii Nauk USSR 60, p.~208--209 (1948) (in Russian).
  \\
  See also~\cite[\textsection 9]{landafshitz}.
 \bibitem{yang}
  C.N. Yang.
  {\it
   Selection rules for the dematerialization of a particle into two photons.
  }
  Physical Review 77, 2, p.~242--245 (1950).
 \bibitem{zgg}
  N. Kanda, R. Abe, T. Fujita, H. Tsuda.
  {\it $Z^0$ decay into two photons. }
  arXiv:1109.0926 (2013).
 \bibitem{higgs-1}
  J. P. Ralston.
  {\it The need to fairly confront spin-1 for the new Higgs-like particle. }
  arXiv:1211.2288v1 (2012).
 \bibitem{bouchiat}
  C. Boushiat, J. Iliopoulos, Ph. Meyer.
  {\it An Anomaly-Free Version of Weinberg's Model. }
  Physics Letters 38B, 7, p.~519--523 (1972).
 \bibitem{landafshitz}
  V. B. Berestetsky, E. M. Lifshitz, L. M. Pitaevsky.
  {\it Quantum Electrodynamics.  }
  Theoretical physics, vol. IV.
  2nd edition.
  Moscow, Nauka. (1980) (in Russian).
 \bibitem{rosenberg}
  L. Rosenberg.
  {\it Electromagnetic interactions of neutrinos. }
  Physical Review 129, 6, p.~2786--2788 (1963).
 \bibitem{jackiw}
  S. B. Treiman, R. Jackiw, D. J. Gross.
  {\it Lectures on current algebra and its applications.  }
  Trans. N. N. Nikolaev and V. A. Novikov.
  Moscow, Atomizdat (1977) (in Russian).
 \end{thebibliography}
\end{document}